\documentclass[aps]{revtex4}
    \def\be{\begin{equation}}
    \def\ee{\end{equation}}
    \def\ba{\begin{eqnarray}}
    \def\ea{\end{eqnarray}}

    \input epsf
    \usepackage{pstcol,graphicx}
\begin{document}
\title{The Impact of Ultraviolet Regularization on the Spectrum of Curvature  
Perturbations During Inflation}

\author{F. Finelli} \email{finelli@iasfbo.inaf.it}
\affiliation{INAF/IASF-BO,
Istituto di Astrofisica Spaziale e Fisica
Cosmica di Bologna \\
via Gobetti 101, I-40129 Bologna - Italy}
\affiliation{INAF/OAB, Osservatorio Astronomico di Bologna,
via Ranzani 1, I-40127 Bologna -
Italy}
\affiliation{INFN, Sezione di Bologna,
Via Irnerio 46, I-40126 Bologna, Italy}

\author{G. Marozzi} \email{marozzi@bo.infn.it}
\author{G. P. Vacca} \email{vacca@bo.infn.it}
\author{G. Venturi} \email{armitage@bo.infn.it}
\affiliation{Dipartimento di Fisica, Universit\`a degli Studi di Bologna
    and I.N.F.N., \\ via Irnerio, 46 -- I-40126 Bologna -- Italy}

\begin{abstract}
Inflationary predictions based on the 
linear theory of cosmological perturbations 
are related to the two point function of a (second quantized) real scalar 
free field during the accelerated stage. Such a two point function is finite, 
in contrast with its coincidence limit, which is divergent due to the 
ultraviolet divergences proper of field theory. We therefore argue that  
predictions of most of the inflationary models do not necessarily need a 
regularization scheme to leading order, i.e. tree level, which is  
required instead for non-linear corrections or calculations involving the 
energy-momentum tensor. 
We also discuss unpleasant features of the "would be" 
regularized spectrum obtained using the traditional 
fourth order adiabatic subtraction.
\end{abstract}

\maketitle


\section{Introduction}\label{one}

Inflationary predictions rely on the heritage of quantum field theory in 
cosmological space-times, pionereed by Parker at the end of the sixties 
\cite{parkersixties}. The nearly scale invariant spectrum of curvature 
perturbations predicted by successful inflationary models is at present 
associated with the concept of amplification of vacuum fluctuations by the 
geometry.

The calculation of the two point function in the coincidence limit and 
of the energy-momentum tensor requires a renormalization scheme, just as 
in ordinary  Minkowski space-time.
Ultraviolet divergences due to fluctuations on arbitrarly short scales 
are common in field theory.
In Minkowski space-time, infinities in a free theory are removed by the
subtraction of the vacuum expectation value of the energy (also called
normal ordering), the physical justification being that these vacuum
contributions are unobservable. In cosmology, according to the 
inflationary paradigm, the large scale fluctuations we see derive from 
quantum vacuum fluctuations. 
However, a prescription similar to the one used in Minkowski 
is used in order to regularize infinities in curved space-times. 
The idea is to subtract from a bare (infinite)  
quadratic quantity its counterpart evaluated for an 
adiabatic change of the geometry.
This procedure is therefore called {\em adiabatic subtraction} 
\cite{adiabaticthesis, adiabatic} and in this way the infinities of field 
theory are removed. 

Although {\em adiabatic subtraction} has been so far only a tool to
remove infinities from divergent quantities, 
it has been argued recently \cite{parker} that a "regularized" spectrum 
may differ substantially from the bare one, 
leading to a rethinking of the inflationary predictions. 
The "regularized" spectrum proposed in \cite{parker} is the bare one minus 
the first two adiabatic terms, i.e. a minimal subtraction scheme for the 
coincidence limit of the two-point function. 

The correlation function of different sets of observations 
- such as the pattern of anisotropies in the cosmic microwave 
background radiation (CMBR) or a catalogue of galaxies - 
is related to the one describing the curvature fluctuation for adiabatic 
initial conditions. 
In the simplest picture, long wavelength curvature perturbations 
remain constant after their exit from the Hubble radius during inflation 
until their re-entry in the Hubble radius during the radiation or matter 
dominated stage. During inflation, a curvature perturbation is proportional 
to the gauge invariant (i.e. with respect to coordinate transformation) 
inflaton fluctuation. 
Therefore, in the simplest picture, the correlation function of the primordial 
curvature fluctuations we derive from observations  
is related to the two-point function 
of inflaton fluctuations {\em taken at different space (or space-time) 
points}. We remark that this two point function of a free scalar field during 
inflation is finite, in contrast with its coincidence limit, and no 
regularization scheme seems necessary to leading order. 
Although Ref. \cite{parker} proposes a minimal subtraction 
scheme for the coincidence limit of the two-point function, 
we show that if one adopts the traditional fourth order adiabatic 
subtraction - which is needed for consistency in the regularization of 
objects evaluated in the coincidence limit - in order to obtain a 
"regularized" spectrum, this latter quantity becomes negative for a range 
of wavelengths. 

Our paper is organized as follows. In sections II we illustrate the 
two-point function of a real minimally coupled scalar field in de Sitter 
space-time. We discuss its coincidence limit and its associated 
adiabatic subtraction scheme in section III. We discuss general 
inflationary models and, in particular, the case of 
power-law inflation in section IV. We then conclude in Section V.

\section{Two Point Function in De Sitter Space-Time}

Let us consider scalar fluctuations with mass $m$ and non minimal coupling 
to the curvature $\xi$  propagating in a cosmological (rigid) space-time, 
whose Fourier modes satisfy the following equation
\be
(a \varphi_k)'' + \Omega^2_k (a \varphi_k) = 0 \,, \quad \quad 
\Omega_k^2 = k^2 + m^2 a^2 + \left( \xi - \frac{1}{6} \right) a^2 R \,,
\ee
where the prime denotes the derivative with respect to the conformal time 
$\eta$. 
Once these fluctuations are quantized in terms of creation and annihilation 
operators ${\hat b}^\dagger \,, {\hat b}$:
\be
\hat \varphi (\eta, {\bf x}) = \frac{1}{(2 \pi)^{3/2}} \int d {\bf k} 
\left[ \varphi_k (\eta) e^{i {\bf k} \cdot {\bf x}} {\hat b}_{\bf k}
+ \varphi^*_k (\eta) e^{- i {\bf k} \cdot {\bf x}} {\hat b}^\dagger_{\bf k}
\right]
\ee
we obtain  the relation between the two point
correlation function of a scalar fluctuation evaluated at different space 
points and the spectrum $P_\varphi$:
\be
\langle \hat \varphi (\eta, {\bf x}) \hat \varphi (\eta, {\bf x'})
\rangle = \int_\ell^{+\infty} \frac{dk}{k} 
\frac{\sin k|{\bf x} - {\bf x'}|}{k \, |{\bf x} - {\bf x'}|}
\frac{k^3 | \varphi_k (\eta)|^2}{2 \pi^2}
\equiv \int_\ell^{+\infty} \frac{dk}{k} 
\frac{\sin k|{\bf x} - {\bf x'}|}{k \, |{\bf x} - {\bf x'}|}
P_\varphi (k,\eta) \,.
\label{twopoint}
\ee
The presence of a non-trivial infrared cut-off $\ell \ne 0$, related to the 
beginning of inflation  \cite{VF,cutoff}, becomes important 
if the spectrum of $\varphi$ is scale invariant or steeper 
(i.e. $d \ln P_\varphi / d \ln k \le 0$) and does not alter the
argument of this paper related to ultraviolet divergencies. 

During the inflationary expansion fluctuations are stretched and 
amplified from initial vacuum fluctuations:
\be
|\varphi_k| \simeq \frac{1}{a \sqrt{2 \Omega_k}} \quad {\rm for} \quad k/a >> 
H \,.
\label{mode_at_uv}
\ee

In order to evaluate the correlation function above (of
Hadamard kind) one should keep in mind that it may be
written as the time coincidence limit of the more general Green's function,
$\langle \hat \varphi (\eta, {\bf x}) \hat \varphi (\eta', {\bf x'}) \rangle$,
which is a solution of the homogeneous equation of motion~\cite{BD} (see also
~\cite{spindel}).
The latter may be computed from the Wightman functions, again solutions of the
homogeneous equations of motion, determined by the usual
prescription which gives a dependence in $\left[(\eta-\eta'-i
  \epsilon)^2-|{\bf x}-{\bf x'}|^2\right]$ and is compatible with the
Minkowsky limit form in suitable cases. Indeed this fact can be seen easily
for a de Sitter space-time~\cite{ratra1985,BDbook}, which is conformally flat.
In the time coincidence limit the Hadamard function
$\langle \hat \varphi (\eta, {\bf x}) \hat \varphi (\eta, {\bf x'})\rangle$
is well defined and depends on $|{\bf x}-{\bf x'}|^2$. 
The above mentioned prescription allows for the convergence of the momentum
integral in the ultraviolet region ( see the integral in Eq. (\ref{twopoint}) 
where the
mode behaviour in (\ref{mode_at_uv}) is considered), which otherwise would be
characterized by an oscillating integrand in the infinite momentum region. 

Although de Sitter space-time does not support Gaussian inflationary
perturbations, it admits an exact solution for test scalar fields
- with arbitrary masses and couplings to the curvature - which for our 
present purposes is similar to that for inflaton fluctuations in a nearly 
de Sitter space-time.
We here just consider a minimal coupling. For an evolution of the scale factor 
given by
\be
a (\eta) = - \frac{1}{H \, \eta} \,,
\quad \quad - \infty < \eta < 0 \,,
\ee
the solution for $\varphi_k$ agreeing with the adiabatic vacuum for
short wavelengths is
\be
\varphi_k = \sqrt{\frac{\pi}{4 H a^3}}
H_\nu^{(1)}\left(\frac{k}{a H}\right) \,, \quad \quad
\nu=\sqrt{\frac{9}{4}-\frac{m^2}{H^2}}\,.
\label{dssolution}
\ee
On considering $m \ne 0$ the infrared divergence is absent and the infrared
cutoff $l$ can be set to zero. 
The de Sitter result for Eq. (\ref{twopoint}) \cite{BD,ratra1985} is given by
\be
\langle \hat \varphi (\eta, {\bf x}) \hat \varphi (\eta, {\bf x'})
\rangle_{\rm dS} = \frac{H^2 \, \sec (\pi \nu)}{16 \pi^2} 
\left(\frac{1}{4} - \nu^2\right) \, _2F_1 \left[ \frac{3}{2} + 
\nu, \frac{3}{2} - \nu \, ; 2 \, ; 1 
- \frac{| {\bf x} - {\bf x'}|^2}{4 \eta^2} \right] \,.
\label{twopointratra}
\ee

We note that this result can be simply obtained from Eq. (\ref{twopoint}) by an
analytic continuation obtained on performing a general integration in
$d$ space dimensions and afterwards taking the $d\to3$ limit.
The analytic structure of the hypergeometric function provides a covering of
the different space-time regions, as is evident if one considers the more 
general correlation
function at different times (in such a case the last argument in the
hypergeometric function is $1+ \left[(\eta-\eta')^2-
|{\bf x}-{\bf x'}|^2\right]/(4\eta \eta')$). In particular in the space-like 
region the
argument ranges from  $1$ (deep inside the Hubble radius) to $- \infty$
(fully outside the Hubble radius), while for time-like regions the argument
is greater than $1$. 
\begin{figure}
\includegraphics[scale=0.54]{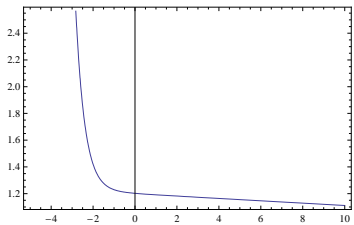}
\caption{$\langle \hat \varphi (\eta, {\bf x}) \hat \varphi (\eta, {\bf x'})
\rangle_{\rm dS}$ in units of $H^2$ for $m^2/H^2=10^{-2}$ as a function of
$\log_{10}{\frac{| {\bf x} - {\bf x'}|^2}{4 \eta^2}}$}
\label{fig1}
\end{figure}
In the space-like region the asymptotic infrared (large distance) behavior
(see Fig.~\ref{fig1}) is given by
\be
\langle \simeq \varphi (\eta, {\bf x}) \hat \varphi (\eta, {\bf x'}) 
\rangle_{\rm dS} \sim 
- H^2 \sec{(\pi \nu)} \frac{2^{2\nu-5}}{\pi^{5/2}}
\frac{\Gamma(\nu)}{\Gamma\left(\nu-\frac{1}{2}\right)}
\left(\frac{|{\bf x} - {\bf x'}|^2}{4 \eta^2}\right)^{\nu-\frac{3}{2}}
\ee

Inflationary models without a classical time dependent
homogeneous background value for the inflaton, such as a test field in a
de Sitter background, were also proposed as models with non-gaussian 
($\chi^2$ distributed) isocurvature perturbations with a blue spectral 
index \cite{MS,LM}. The experimentally interesting
quantities are related to correlation functions of such objects and at a
quantum level in such a simple free case one has from Wick's theorem
\be
\langle \hat \varphi^2 (\eta, {\bf x}) \hat \varphi^2 (\eta, {\bf x'})
\rangle -\left( \langle \hat \varphi^2 (\eta, {\bf x}) \rangle \right)^2=
2 \left( \langle \hat \varphi (\eta, {\bf x}) \hat \varphi (\eta, {\bf
    x'})\rangle \right)^2 
\label{zeroclassical}
\ee
Therefore, in this case the subtraction 
(which implements a trivial renormalization) is
included in the definition of a connected Green's function.

\section{Regularization of the Coincidence Limit of the Two Point Function}

Let us try to understand why one may be tempted to introduce a deformation
of the spectrum although there is no necessity to do it. 
When one is interested in the vacuum expectation value of composite quantum 
operators such us the energy-momentum tensor, the appearance of divergences of
ultraviolet origin calls for a renormalization procedure which normally 
consists of two steps:
regularization (in order to work with finite quantities) and subtraction (to
remove the divergent pieces).
The simplest of such operators is $\varphi^2(x)$. In order to
have a regularized quantity for its vacuum average one may consider 
point-splitting \cite{BDbook}, also other approaches, such as the dimensional 
regularization, are often considered.
As stated before a possible way of removing the infinities is given by the 
adiabatic subtraction \cite{adiabaticthesis, adiabatic}. To be in agreement
with the renormalized effective action results \cite{BDbook}, 
which also determine the
anomalous contribution to the trace of the energy momentum tensor averaged
over the vacuum, a fourth order adiabatic subtraction is needed.
In fact, this is also needed to remove the ultraviolet divergence for the 
energy-momentum tensor in a general space-time (as can be seen, in the case 
of inflationary models with scalar metric fluctuations, from the
general adiabatic expansion described in \cite{Marozzi}).

Let us consider the coincidence limit of Eq. (\ref{twopoint})
\be
\langle \hat \varphi^2 (\eta, {\bf x})
\rangle = \int_\ell^{+\infty} \frac{dk}{k}
\frac{k^3 | \varphi_k (\eta)|^2}{2 \pi^2}
\equiv \int_\ell^{+\infty} \frac{dk}{k} P_\varphi (k,\eta) \,.
\label{Spectra}
\ee
This quantity is not defined since it is divergent ($| {\bf x} - {\bf
  x'}|^2\ne 0$ can be seen as a regulator in the point-splitt version).
Nevertheless, on using the adiabatic subtraction, the renormalized value 
can be obtained in different ways:
one can first perform a dimensional  regularization of the divergent integral
and then subtract the fourth order adiabatic value, safely removing the
regularization afterwards to obtain the finite renormalized expectation value:
\begin{eqnarray}
\langle \hat \varphi^2\rangle_{\rm REN} &=& \lim_{d\rightarrow 3}
\left(\langle\varphi^2\rangle -\langle\varphi^2\rangle_{(4)} \right) 
\nonumber \\
&=& \lim_{d\rightarrow 3} \left\{ \frac{1}{(2 \pi)^d} 
\frac{2 \pi^{d/2}}{\Gamma(d/2)}
\left[ \int_{\ell}^{\infty} dk \, k^{d-1} \,
|\varphi_k|^2 - \int_{\ell}^{\infty} dk \, k^{d-1} \,
|\varphi_k^{(4)}|^2 \right]\right\} \,.
\label{dimension_regularization}
\end{eqnarray}
Or else one may subtract the fourth adiabatic order before performing the 
integrals without using an additional regularization procedure:
\be
\langle \hat \varphi^2\rangle_{\rm REN}
=\frac{1}{2 \pi^2} \int_{\ell}^{\infty} dk \, 
k^2  \,\left[ |\varphi_k|^2-|\varphi_k^{(4)}|^2\right] \,.
\label{adiabatic_subtraction}
\ee

These two ways of proceeding give, with the presence of a non zero infrared 
cut-off $l$, always the same result. Without a non zero infrared cut-off there 
are some interesting exceptions, for the particular case of a de Sitter 
space-time with $m=\xi=0$, for instance, 
those two methods fail to reproduce the right value 
for $\langle\varphi^2\rangle_{\rm REN}$ (given by \cite{VF,cutoff,AF}) while
for the renormalized value of the energy-momentum tensor, which has 
no infrared divergence, only the second method is applicable, on proceeding as 
described in \cite{FMVV_GW}, obtaining the correct Allen-Folacci value
\cite{AF}.

We now go back to our particular case of a de Sitter space-time, the  
$\varphi_k^{(4)}$ expansion up to the fourth adiabatic order is given by
(using the result of \cite{FMVV}):
\begin{eqnarray}
|\varphi_k^{(4)}| = \frac{1}{a \sqrt{2 \Sigma_k}}
&\Bigl\{&
1+\frac{a^2 R}{24} \frac{1}{\Sigma_k^2}+\frac{5}{32} \frac{a^4 R^2}{36}
 \frac{1}{\Sigma_k^4} \nonumber \\
 & & +\frac{1}{8}\frac{1}{\Sigma_k^4}
2 \left( a a^{''} + a^{' \, 2} \right) \left(m^2-\frac{R}{6}
 \right) \nonumber \\
 & &  -\frac{5}{16}\frac{1}{\Sigma_k^6}\left[a^2 a^{' \, 2}m^4- a^2 
a^{'2}m^2
 \frac{R}{3}\right] \nonumber \\
 & & +\frac{3}{64}\frac{1}{\Sigma_k^6} R 
\left( a a^{''} + a^{' \, 2} \right) m^2-
\frac{65}{64} \frac{1}{\Sigma_k^8}
a^2 \frac{R}{6} a^{'\,2}m^4+\frac{5}{32}\epsilon_{2*}^2-
\frac{1}{4}\epsilon_{4*}   \Bigr\}
\label{fourth_bis}
\end{eqnarray}
with 
\begin{eqnarray}
    \Sigma_k=(k^2 + a^2 m^2)^{1/2} \quad , \quad
    \epsilon_{2*}=-\frac{1}{2}\frac{\Sigma_k^{''}}{\Sigma_k^3}+\frac{3}{4}
    \frac{\Sigma_k^{'2}}{\Sigma_k^4} \quad , \quad
    \epsilon_{4*} =
    \frac{1}{4}\frac{\Sigma_k^{'}}{\Sigma_k^3}\epsilon_2^{'}-
    \frac{1}{4}\frac{1}{\Sigma_k^2}\epsilon_2^{''}
\label{eps24}
    \end{eqnarray}

Let us note that the adiabatic terms are defined over the whole $k$ range. 

Although the subtraction of the adiabatic terms are required 
in order to obtain a finite answer for quantities integrated over (almost) 
the whole $k$ range, it is worth investigating whether the 
subtracted adiabatic terms $k$ by $k$ do spoil 
$P_\varphi$, which should be positive by physical reasons and
whose shape is in agreement with observations and constitutes the major
success of inflation. We shall therefore investigate 
\be
{\cal P}_\varphi^{(n)} (k,\eta) = P_\varphi (k,\eta) 
- P_\varphi^{(n)} (k,\eta) \,,
\label{Pren}
\ee
which is the most immediate "regularized" spectrum, given 
the adiabatic subtraction scheme illustrated in 
Eq. (\ref{adiabatic_subtraction}). 

In \cite{parker} it has been chosen 
to make the subtraction only up to second order since, 
as is clear from Eqs.(\ref{fourth_bis}, 
\ref{eps24}) for the particular case of a de Sitter space-time, this is 
enough to remove all the infinities in Eq. (\ref{adiabatic_subtraction}).
But, as we state at the begining of this section, although 
for $\langle \hat \varphi^2\rangle$ the
ultraviolet divergence are present only in the first two adiabatic orders,
the fourth order adiabatic subtraction  may be needed 
in order to have a self-consistent scheme.

For a de Sitter space time Eq. (\ref{Pren}) gives:
\begin{eqnarray}
{\cal P}_\varphi^{(4) \, dS} (k,\eta)
&=&   \frac{k^3}{4 \pi^2} \Biggl\{\frac{\pi}{2 H a^3} 
\left|H_\nu^{(1)}\left(\frac{k}{a H}\right)\right|^2 -\frac{1}{a^2 \Sigma_k}
-\left[ -\frac{5}{8}\frac{a^4 m^4 H^2}{\Sigma_k^7}+\frac{3}{4}
\frac{a^2 m^2 H^2}{\Sigma_k^5}+\frac{H^2}{\Sigma_k^3}\right]+
 \nonumber \\ & & 
-\left[ \frac{1155}{128}\frac{a^{10} m^8 H^4}{\Sigma_k^{13}}-\frac{693}{32}
\frac{a^{8} m^6 H^4}{\Sigma_k^{11}}+\frac{385}{32}
\frac{a^6 m^4 H^4}{\Sigma_k^9}+\frac{5}{2}\frac{a^4 m^2
  H^4}{\Sigma_k^7}\right]
\Biggr\} \,.
\label{deSitter_spectra}
\end{eqnarray}
A key point in discussing Eq. (\ref{deSitter_spectra}) 
is the different time evolution of the mode $\varphi_k$ and its
adiabatic approximation. A nearly scale invariant spectrum
is given by a minimally coupled scalar field with $m << H$,
whose Fourier mode $\varphi_k$ is almost frozen in time after
it has crossed the Hubble radius, although starting from the adiabatic
vacuum when it is well within the latter. As is clear from Eq.
(\ref{fourth_bis}), adiabatic terms decay at least as
${\cal O} (1/a^2)$ for wavelengths much larger than the Compton one
$k/a << m ( << H )$. Because of this differing behaviour in time, the infrared
part of the spectrum is left unchanged by the subtraction in any case.
Indeed, for scales well outside the Hubble radius $k << a H$:
\be
\frac{k^3}{2 \pi^2} \frac{\pi}{4 H a^3}
\left|H_\nu^{(1)}\left(\frac{k}{a H}\right)\right|^2 \simeq 
\frac{k^3}{2 \pi^2} \frac{\pi}{4 H} \frac{\Gamma^2 (\nu)}{\pi^2 a^{3-2\nu}} 
\left( \frac{2 H}{k} \right)^{2 \nu} 
\ee
which shows how the bare part weakly depends on time 
(${\cal O} (a^{-\frac{2 m^2}{3 H^2}})$ for $m << H$).

\begin{figure}
\begin{tabular}{cc}
\includegraphics[scale=0.54]{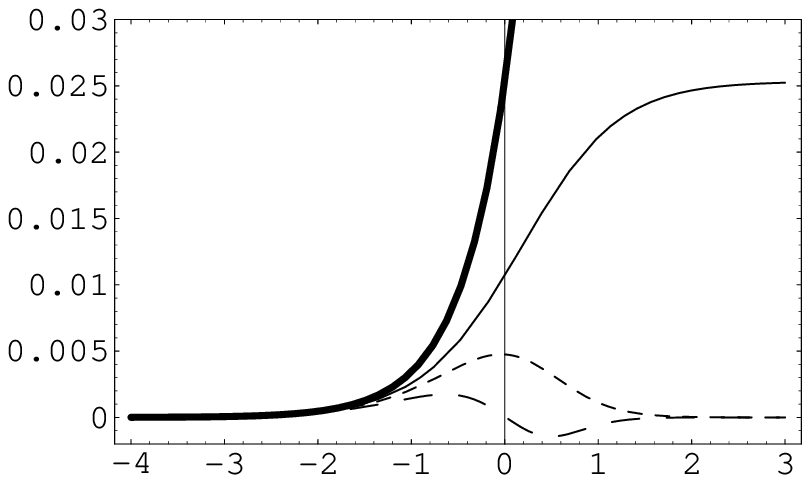}
\includegraphics[scale=0.54]{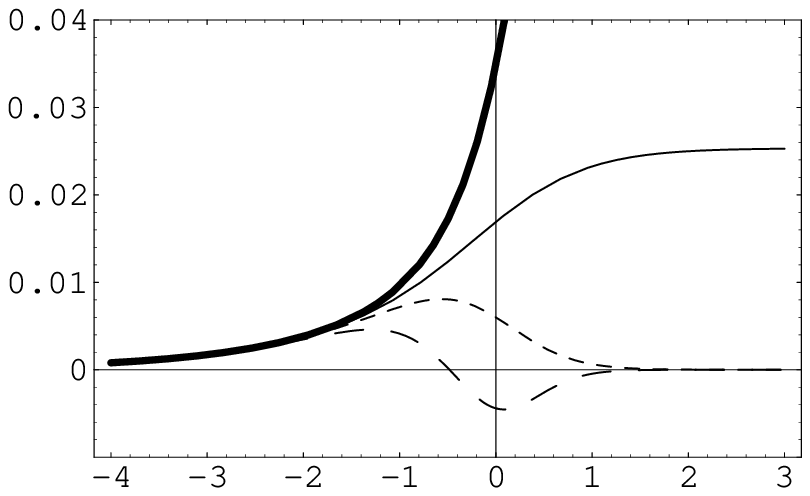}
\includegraphics[scale=0.54]{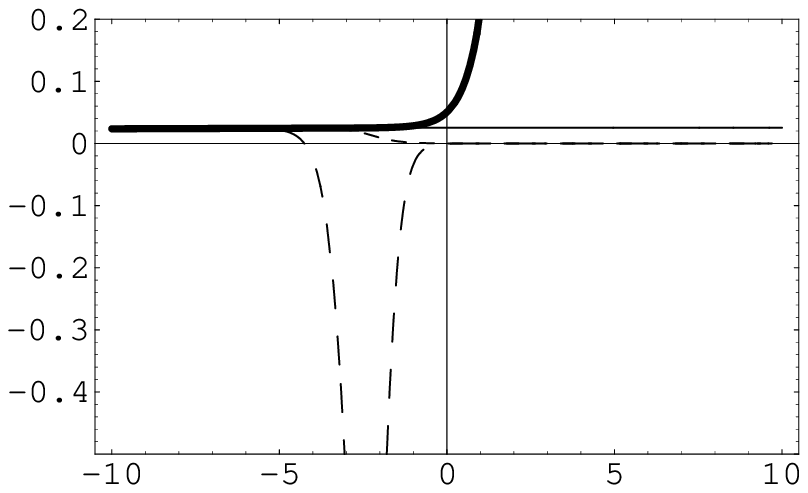}
\includegraphics[scale=0.54]{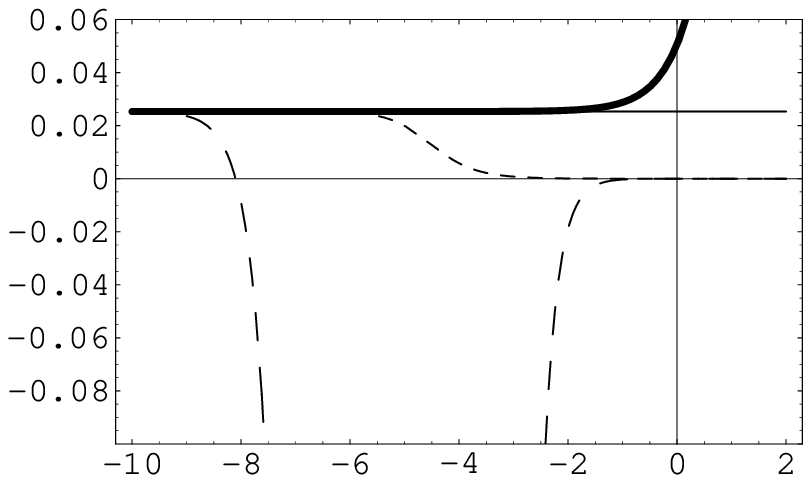}
\end{tabular}
\caption{Plots of the following power spectra in units of $H^2$ vs 
$\ln (k/a H)$ in the de Sitter case: 
$P_\varphi$ (thick line), ${\cal P}_\varphi^{(0)}$ (solid line), 
${\cal P}_\varphi^{(2)}$ (short-dashed line), 
${\cal P}_\varphi^{(4)}$ (long-dashed line). 
From the left to the right plot, $m^2/H^2 = 2 \,, 1 \,, 10^{-2} \,, 
10^{-4}$. Note how the fourth order adiabatic spectrum 
${\cal P}_\varphi^{(4)}$ becomes negative for any value of $m$, even if for 
different values of $k$.}
\label{fig2}
\end{figure}

\begin{figure}
\begin{tabular}{cc}
\includegraphics[scale=0.54]{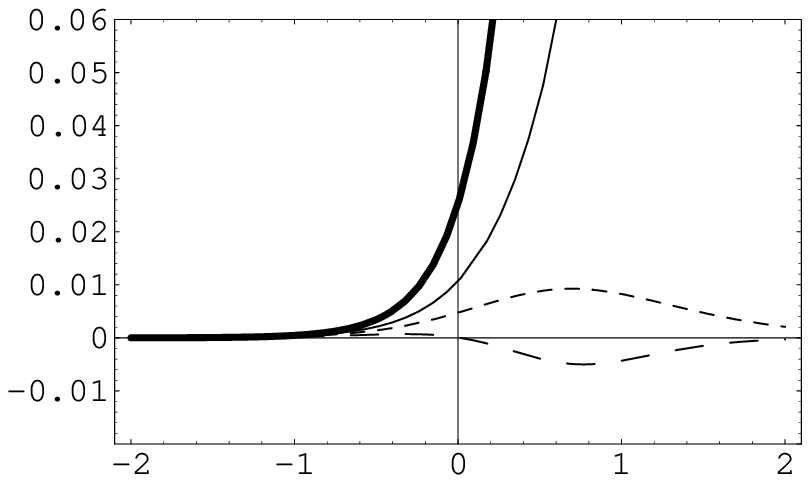}
\includegraphics[scale=0.54]{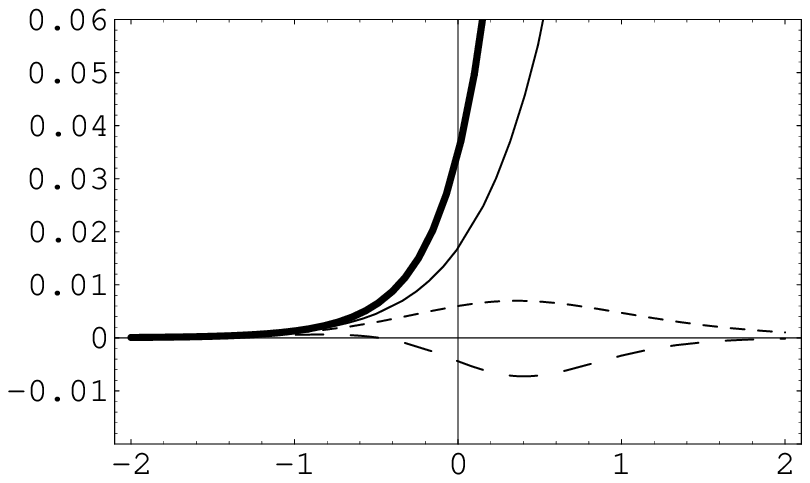}
\includegraphics[scale=0.54]{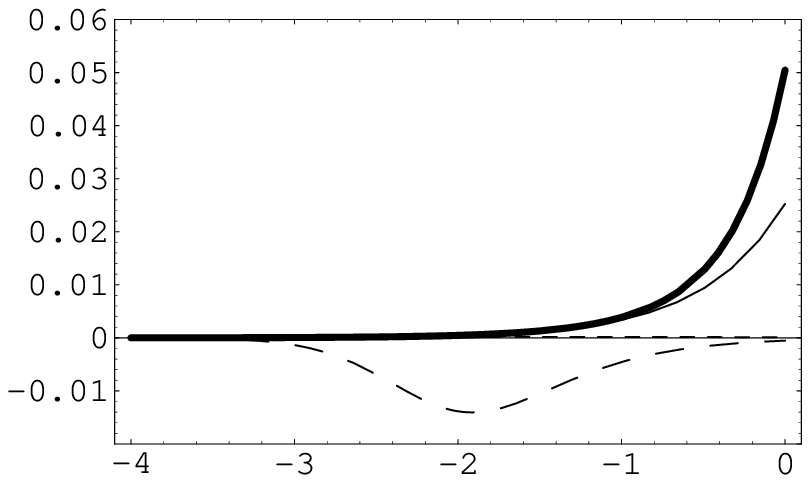}
\includegraphics[scale=0.54]{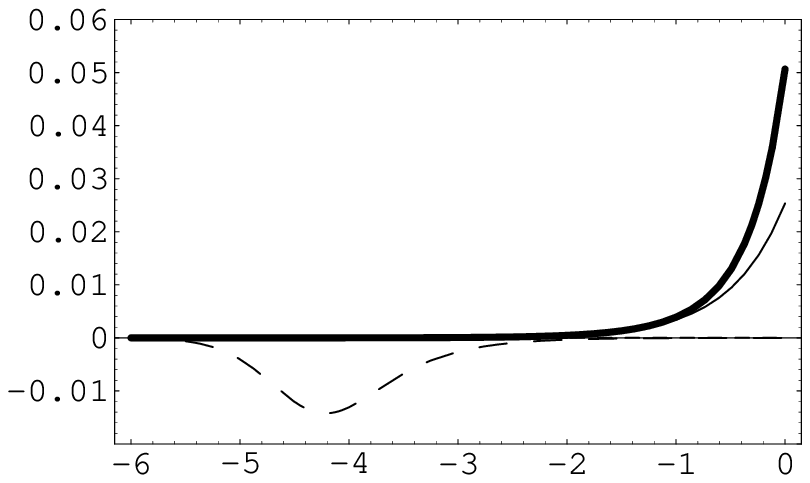}
\end{tabular}
\caption{Same as Fig. 2, but for $k^2 |\varphi_k|^2/a^2$ (in units of $H^4$), 
which is quartically divergent.}
\label{fig3}
\end{figure}

Fig. \ref{fig2} shows the bare PS and its regularized forms for various values 
of the mass $m$. Note that the fourth order regularized 
PS ${\cal P}_\varphi^{(4)}$ becomes always negative for a different range 
of wavelengths for any value of the mass $m$. Fig. 2 shows the same for the 
laplacian term on making an adiabatic expansion only for $|\varphi|$.

\section{Inflationary Examples}\label{four}

Inflationary models driven by a classical zero mode support 
Gaussian inflationary fluctuations. Scalar metric fluctuations should be 
taken into account with field fluctuations, leading to 
gauge-invariant curvature perturbations 
\be
R = \frac{H}{\dot \phi} Q_\varphi
\ee
with $Q_\varphi$ being the Mukhanov variable.
In the Uniform Curvature Gauge this coincide with the inflaton fluctuation and
satisfies:
\be
\ddot Q_{\varphi \, k}+3 H \dot Q_{\varphi \, k}+
\left [\frac{k^2}{a^2} + V_{\varphi \, \varphi} 
+ 2 \frac{\dot H}{H} \left(3 H -
\frac{\dot H}{H} +
2 \frac{\ddot{\phi}}{\dot{\phi}} \right) \right] Q_{\varphi \, k} = 0 \,.
\label{mukhanov}
\ee

Power-law inflation \cite{LucMat} with 
$V (\phi) \propto e^{-\lambda \frac{\phi}{M_{\rm pl}}}$
is one of the few cases for which solutions for inflaton fluctuations 
are known on the inflationary trajectory for which $a (t) \sim t^p$ with 
$p > 1$. On the above trajectory Eq. (\ref{mukhanov}) becomes an equation 
for a massless minimally coupled scalar field (the same as the 
polarization amplitude for gravitational waves). The exact solution is:
\be
\varphi_k = \sqrt{-\frac{\pi \eta}{4 a^2}}
\, H_\alpha^{(1)}\left(-k\eta\right) \,, \quad \quad
\alpha=\frac{3}{2}+\frac{1}{p-1}
\label{plsolution}
\ee

Although the above exact solutions are very similar to the ones in Eq. (6) 
for the de Sitter case, the two point correlation function for power-law 
inflation depends on the infrared cut-off since the spectrum for $Q_k$ is 
red-tilted \cite{PF}. 
Such an infrared divergence in the Fourier integral 
is general for all the inflationary models predicting a red-tilted 
spectrum of cosmological perturbations and of course persists in the 
coincidence limit. The final formula for the two point function will 
therefore be dependent on the infrared cut-off (which may be related to 
the beginning of inflation): in this case the power spectrum will not be 
simply the one obtained from Eq. (\ref{plsolution}), but it will be changed 
for very small $k$. 
Such a correction may be relevant in the case of a short inflationary stage 
and/or a very smooth beginning of inflation. In some extreme cases, such 
infrared modifications required by theory may lead to observable effects 
\cite{CPKL}.

\begin{figure}
\begin{tabular}{cc}
\includegraphics[scale=0.54]{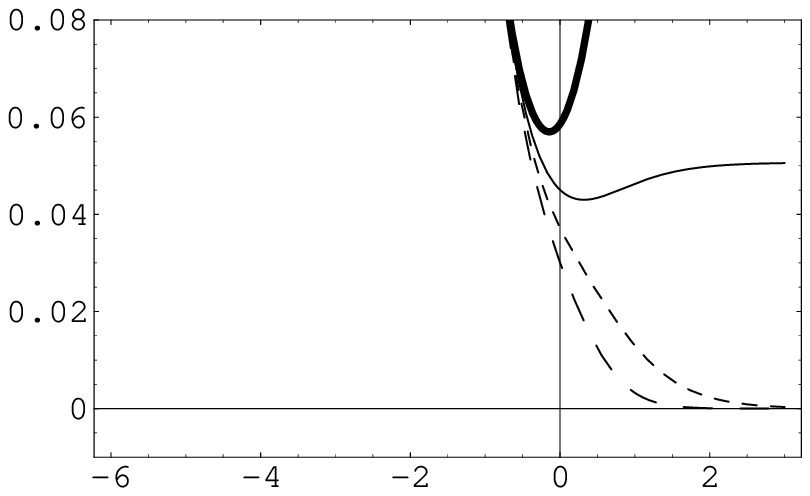}
\includegraphics[scale=0.54]{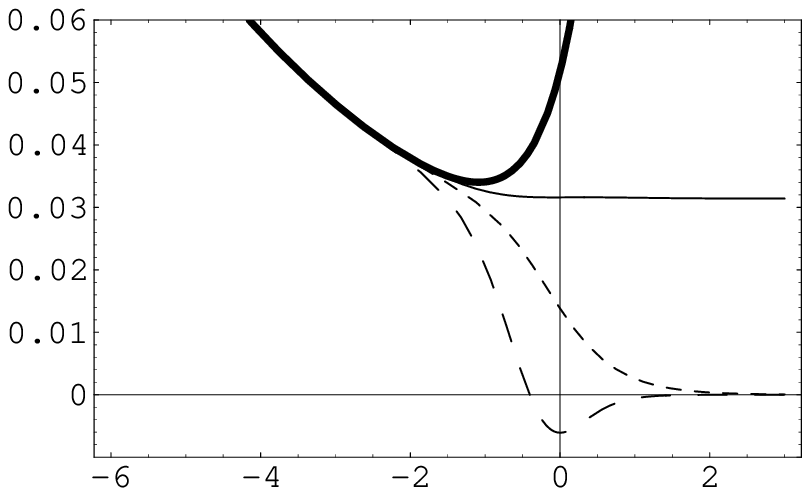}
\includegraphics[scale=0.54]{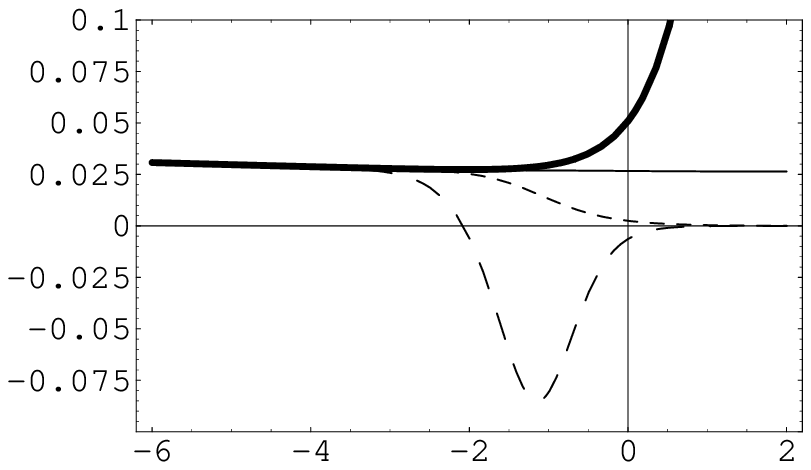}
\includegraphics[scale=0.54]{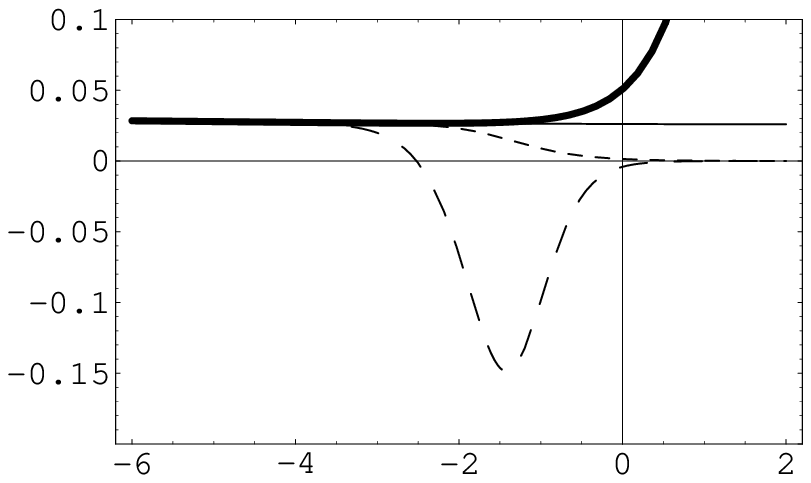}
\end{tabular}
\caption{Plots of the following power spectra in units of $H^2$ vs 
$\ln (k/a H)$ in a power-law model of inflation: 
$P_Q$ (thick line), ${\cal P}_Q^{(0)}$ (solid line), 
${\cal P}_Q^{(2)}$ (short-dashed line), 
${\cal P}_Q^{(4)}$ (long-dashed line). 
From the left to the right plot, $p=2, \, 10, \, 60, \, 100$. Note how the 
fourth order adiabatic spectrum 
${\cal P}_Q^{(4)}$ becomes negative for the larger values of $p$, 
even if for different values of $k$.}
\label{figPL}
\end{figure}

Also in such a simple case of inflation, ${\cal P}_Q^{(4)} (k)$ can 
become negative. 
Proceeding as in the de Sitter case and using the result of 
\cite{Marozzi} for the adiabatic part, ${\cal P}_Q^{(4)} (k)$ is shown 
in Fig. \ref{figPL} for different values of $p$. 
The adiabatic expansion for the gauge-invariant field fluctuation 
for a general potential $V(\phi)$ is given in \cite{Marozzi}, and 
therefore any model can be studied at tree-level in an analogous way.

\section{Conclusions}

We have discussed the necessity of introducing a deformation in the bare 
power spectrum of inflationary fluctuations due to 
regularization/renormalization prescriptions. 
We have shown that there is no necessity to concern oneself about ultraviolet 
divergencies for all inflationary models as long as the fluctuations 
are studied for free fields to lowest order: the two-point function (taken 
at different points) of fluctuations is finite. 
This conclusion apples also to inflationary models without a time dependent 
homogeneous value, as shown in Eq. (\ref{zeroclassical}). 
In contrast, the fluctuation two point function requires a prescription 
in the infrared for those red-tilted spectra which are in agreement with 
observations. 
To lowest order there is no ultraviolet renormalization to be invoked 
and we have shown that the deformation of the spectrum proposed in
\cite{parker}, if extended to a four order subtraction,  
may lead to additional problems, such as "regularized" spectra 
which are not positive in a certain range of $k$.  

Beyond linear order, first corrections to
correlation functions (of linear quantities) appear instead at quartic order 
in the fields. There can be different approaches which start from Einstein
equations~\cite{FMVV2006} or from the associated
effective action for the fluctuations~\cite{weinberg,sloth}.
In both cases it appears clear that the inclusion of perturbative corrections
to the correlation functions requires taking into account
a proper renormalization procedure, which removes the ultraviolet divergences
arising from loop diagrams, and the consequent redefinition of the couplings.

\section{Acknowledgements}

We wish to thank L. Parker for kind correspondence and useful 
comments on the manuscript.

\end{document}